# Black Holes, Galaxy Formation and the Large-scale Structure of the Universe [1]

Michael J. West

Department of Astronomy & Physics

Saint Mary's University

Halifax, Nova Scotia B3H 3C3, Canada

west@sisyphus.stmarys.ca

## ABSTRACT

A new model for the formation of active galaxies is described. A key feature of this model is the idea that the birth of black holes in the centers of supergiant galaxies is strongly influenced by the large-scale distribution of matter in the universe. This conjecture can successfully explain two observed phenomena: the alignment of the radio, optical and infrared axes of high-redshift radio galaxies, and the alignment of present-day cD galaxies with their environs. Other recent observational tests of this theory are also discussed.

## 1. Introduction: A Tale of Two Alignment Effects

Black holes are widely believed to be the central engines that power active galaxies such as quasars and radio galaxies. Yet how and when such black holes formed remains one of the great outstanding problems in astrophysics.

In this paper, a truly remarkable link is postulated between the formation of black holes, massive galaxies and the large-scale structure of the universe. Motivation for this work came from two observed phenomena discussed below.

### 1.1. Observations of high-redshift radio galaxies

Considerable insight has been gained in recent years into the nature of distant galaxies. Because of the finite speed of light, distant galaxies are seen as they were in the past, which allows astronomers to directly probe the earliest stages of galaxy formation. Radio observations have proven to be a particularly effective method of finding high-redshift galaxies. At present the most distant known radio galaxies and quasars are at redshifts $z \geq 4$, which correspond to epochs when the universe was less than 10% its present age.

Optical observations of powerful high-redshift radio galaxies have revealed that they are often very elongated and clumpy in appearance. An example is shown in Figure 1. This appearance is distinctly different from that of most galaxies in the present-day universe. The high incidence of double- and multiple-component radio galaxies with separations of only a few thousand light-years

---

Figure 1. Hubble Space Telescope image of the distant radio galaxy 3C 324 at a redshift $z = 1.21$. Note the very clumpy and elongated appearance of this galaxy, which is characteristic of many powerful radio galaxies at high redshifts.

Figure 2. The alignment effect in high-redshift radio galaxies. Shown here is the frequency distribution of the difference between the radio and optical axis orientations of a sample of powerful, high-redshift radio galaxies. If the radio and optical axes were uncorrelated, then a uniform distribution between $0°$ and $90°$ would be expected. Instead, a strong tendency is seen for the radio and optical axes to share the same orientation; this alignment tendency is significant at greater than the 99.9% confidence level.

(e.g., 3C 65, 114, 265, 324, 356 and 368) strongly suggests that these systems are in the process of merging.

Yet perhaps the most intriguing property of powerful, high-redshift radio galaxies is the observed alignment of their extended optical and infrared structures with their radio jets (e.g., Chambers, Miley & van Breugel 1987; McCarthy et al. 1987; Rigler et al. 1992; Dunlop & Peacock 1993). This alignment effect is illustrated in Figure 2. This phenomenon is puzzling, since the optical and infrared emission presumably trace the distribution of stellar light (and hence the mass distribution) in these galaxies, whereas the orientation of the radio jet axis is likely determined by physical processes on much smaller scales associated with the black hole central engine. A number of different mechanisms have been proposed for the origin of this alignment effect, however to date none has proven entirely satisfactory.

### 1.2. Observations of low-redshift cD galaxies

There is a second alignment effect, seemingly unrelated to the first, which involves cD galaxies.

cD galaxies are among the most enigmatic objects in the sky. Found exclusively in the centers of galaxy clusters, cDs are the most massive galaxies in the present-day universe, and the most luminous objects whose light is purely stellar in origin. A typical cD galaxy is shown in Figure 3.

*Figure 3. Optical image of a typical cD galaxy, located in the galaxy cluster Abell 3656.*

These supergiant galaxies are generally quite elongated in appearance, and observational evidence suggests that they are predominantly prolate in shape ($l_1 \gg l_2 \gg l_3$, where $l_i$ are the principal axes of the galaxy mass distribution; Porter, Schneider & Hoessel 1991; Ryden, Lauer & Postman 1993).

One of the most striking features of the large-scale distribution of luminous matter in the universe is its filamentary appearance, with long, quasi-linear arrangements of galaxies extending over hundreds of millions of light-years (see Figure 4). Interestingly, cD galaxies exhibit a strong propensity for their principal axes to align with the surrounding filamentary matter distribution (e.g., Binggeli 1982; Lambas, Groth & Peebles 1988; van Kampen & Rhee 1990; West 1994; West

Figure 4. *A portion of the observed large-scale galaxy distribution. Each dot represents a galaxy as seen in projection on the plane of the sky. Filamentary features can be seen extending over many millions of light-years in length.*

Figure 5. *The alignment effect in low-redshift cD galaxies. Shown here is the frequency distribution of the difference between the orientation of the cD galaxy principal axis and the direction defined by the surrounding large-scale filamentary matter distribution. The significance of the alignment effect exceeds the 99.9% confidence level.*

& Schombert 1996). This alignment effect is shown in Figure 5. It is certainly remarkable that the structure of cD galaxies should somehow know about the distribution of matter on scales of tens or hundreds of millions of light-years. No other type of galaxy exhibits this sort of alignment effect, which suggests that cD galaxies must be the result of a unique formation history unlike that of other galaxies.

### 1.3. Are the two alignment effects related?

To date, the two aligment effects described above have been assumed to be separate and unrelated phenomena. However, as the following sections will endeavour to show, a compelling argument can be made that the two alignment effects are in actuality *related* phenomena that share a common origin.

## 2. cD Galaxy Formation via Anisotropic Mergers

In most currently popular models for the evolution of structure in the universe, galaxy formation is envisioned as proceeding in a hierarchical manner. Supergiant galaxies like cDs would likely have formed by a series of mergers of smaller galaxies.

Given the filamentary topology of the large-scale matter distribution, it seems rather unlikely that such mergers would have occurred in a random, isotropic manner. Instead, the mergers which give rise to cD galaxies at the centers of clusters would most likely occur along preferred directions as material infalls anisotropically along filaments. This special formation history produces cD galaxies that are quite *prolate* in shape and that have preferred orientations built into them which reflect the surrounding filamentary pattern of the large-scale matter distribution, even at early epochs (West 1994). State-of-the-art numerical simulations confirm this claim (see Figure 6). This presumably is the origin of the alignment effect seen in Figure 5.

## 3. The Connection Between Powerful, High-redshift Radio Galaxies and Low-redshift cD Galaxies

A number of pieces of evidence suggest that powerful radio galaxies and quasars at high redshifts were the precursors of today's cD galaxies. To begin with, observational studies have shown that high-redshift radio galaxies and quasars generally reside in dense environments comparable to those of present-day cD galaxies (e.g., Yee & Green 1987; Yates, Miller & Peacock 1989; Ellingson et al. 1991). Furthermore, it is well established that the most powerful radio galaxies in the low-redshift universe are often cD galaxies, and thus it seems likely that powerful radio sources at earlier epochs were also associated with cD galaxies or their progenitors. Additional support for this idea comes from order of magnitude estimates which suggest that the space density of high-redshift radio galaxies and quasars may have been comparable to the present-day space density of cD galaxies (West 1994). It is worth noting too that the cD galaxy in the simulation shown in Figure 6 has the same sort of clumpy and elongated appearance at early epochs that is characteristic of powerful, high-redshift radio galaxies (compare Figures 1 and 6).

These and other arguments point strongly to an evolutionary link between powerful, high-redshift radio galaxies and low-redshift cD galaxies. Having argued earlier that highly anisotropic mergers will produce prolate cDs with preferred orientations, the effect of such a formation history on the genesis of radio sources at high redshifts is considered in the following section.

*Figure 6. Computer simulation of cD galaxy formation (kindly provided by Ray Carlberg). This has been extracted from a larger million-particle simulation of a universe dominated by cold dark matter (see Carlberg 1994). Each box is approximately 13 million light-years across. Time sequence is denoted by the redshift z. Only those particles that end up in the final cD galaxy are shown here. Note the highly anisotropic nature of the merger process, and how this is reflected in the shape and orientation of the galaxy at both high and low redshifts.*

## 4. What Determines the Jet Orientation in Powerful Radio Galaxies?

For several decades now, the the standard model for active galaxies has envisioned a massive compact object, most probably a black hole of order $10^7 - 10^9$ solar masses, which acts as a central engine in the production of radio emission (e.g., Lynden-Bell 1969; Zel'dovich & Novikov 1971; Blandford & Rees 1974; Begelman, Blandford & Rees 1984). According to this picture, the radio jets are produced by twin beams of relativistic material that are ejected along the black hole spin axis.

At early epochs, frequent mergers of small proto-galactic systems would have provided an abundant source of gas-rich material in burgeoning young cD galaxies (see Figure 6). Cold gas and dust falling into the prolate gravitational potential of a developing cD would quickly settle into a thin rotating disk as dissipative processes damp out random motions. It is straightforward to show from dynamical considerations that a rotating gaseous disk in a prolate gravitational potential will precess until its angular momentum vector is aligned with the principal axis of the galaxy's mass distribution (Kahn & Woltjer 1959; Gunn 1979). Angular momentum loss during this disk precession results in an inward flow of gas towards the center of the galaxy, where its accumulation provides a plentiful source of material for the creation and fuelling of a black hole (Rees 1978; Gunn 1979).

In this scenario of black hole formation, one would expect the black hole spin axis to share the same orientation as the angular momentum vector of the fuelling accretion disk, which is ultimately determined by the intrinsic shape and orientation of the host cD galaxy (e.g., Tubbs 1980; Tohline et al. 1981; Steiman-Cameron & Durisen 1988). Recent Hubble Space Telescope observations have provided quite convincing evidence of accretion disks associated with black holes in the centres of two galaxies (Jaffe et al. 1993; Harms et al. 1994); as expected, in both cases, the radio jet axis is perpendicular to the fuelling accretion disk.

Because cD galaxies are prolate in shape as a result of the anisotropic merger process by which they formed, the black hole spin axis, and hence radio jets, will tend to align with the major axes of these galaxies, even at high redshifts. Hence, the anisotropic merger model proposed here for the origin of the alignment effect in cD galaxies also provides a natural mechanism to explain the alignment of the radio, optical and infrared axes in high-redshift radio galaxies.

## 5. Low-redshift Counterparts to High-redshift Radio Galaxies

It is interesting to note that some of the most powerful radio galaxies at low redshifts exhibit *both* alignment effects, which suggests that they may indeed have formed in the manner proposed here.

A good example is the cD/radio galaxy NGC 708, which is located in the galaxy cluster Abell 262 (see Figure 7) and forms part of the well-known Perseus-Pisces filament. NGC 708 is known to have a gas/dust lane that is oriented nearly perpendicular to its radio source axis. This radio axis shares a similar orientation as the galaxy principal axis, which in turn is aligned with the surrounding large-scale filamentary feature seen in Figure 7.

Another example is provided by Hercules A (3C 348), an extremely powerful low-redshift ($z = 0.154$) radio source. Hercules A is a very elongated cD galaxy which resides in a cluster of galaxies (Yates, Miller & Peacock 1989). As Figure 8 shows, Hercules A exhibits the same alignment effect seen in its high-redshift cousins; its radio and optical axes share the same orientation, which indicates that Hercules A is most probably prolate in shape. Although 3C 348 is too far away for the surrounding filamentary galaxy distribution to be seen in existing surveys, Abell clusters can serve as beacons for mapping the surrounding large-scale matter distribution (in Figure 7, for example, the Abell clusters trace the same large-scale structure as the galaxies, albeit more sparsely). As Figure 8 shows, the radio and optical axes of Hercules A appear to "point" along an unseen filament traced by its nearest companion clusters, Abell 2210 and Abell 2204.

Still another example is Cygnus A (3C 405), which is by far the most powerful radio source in the low-redshift universe. Its proximity affords a unique opportunity to study a nearby galaxy whose properties are likely to be quite similar to those of powerful radio galaxies at high redshifts. As shown by West (1994), the orientations of the radio and optical structures of Cygnus A, and the galaxy cluster in which it resides, are remarkably similar. Most importantly, Cygnus A's radio jet clearly "points" towards its nearest neighbouring cluster, Abell 2319, as can be seen in Figure 9. This strongly suggests that Cygnus A has formed via a process of anisotropic mergers that have influenced not only the orientation of the galaxy, but also the orientation of its radio emission – and by implication, the formation of its central black hole.

Figure 7. Comparison of the radio, optical and dust-lane orientations in the cD galaxy NGC 708. Also plotted here is the large-scale distribution galaxies in the same region of the sky. Circles denote Abell clusters of galaxies. Note the tendency for the radio and optical axes of NGC 708 to point along the local large-scale filament.

*Figure 8. The alignment effect(s) in Hercules A (3C 348). The lower panels show optical (left) and radio (right) images of Hercules A, while the top panels shows Hercules A in relation to its large-scale surroundings. Note the tendency for the radio and optical axes to point towards the neighbouring Abell clusters.*

Figure 9. The alignment effect(s) in Cygnus A (3C 405). The top panel shows Cygnus A in relation to its nearest neighbouring cluster of galaxies, Abell 2319. The two bottom panels show the optical (left) and radio (right) structure of Cygnus A. Note the clear alignment of the radio jets with the direction towards the neighbouring cluster.

## 6. Predictions of the Anisotropic Merger Model

Any successful theory should strive not only to explain existing observations, but also to make specific predictions that can be tested. In this spirit, West (1994) proposed a number of tests of the anisotropic merger model described here. Recent observational work has confirmed several of these predictions:

(1) *Anisotropic distribution of companion galaxies around high-redshift radio galaxies.* Given the proposed anisotropic distribution of material around proto-cD galaxies, companion galaxies around high-redshift radio galaxies and quasars should also be anistropically distributed, with a strong tendency for such objects to be found along the direction defined by the radio axis. This means not only clumps associated with the radio galaxy itself, but also other companion galaxies at greater separations. This important prediction of the anisotropic merger model has recently been confirmed by Röttgering, West, Miley & Chambers (1995).

(2) *Formation of clusters of galaxies via anisotropic mergers of subclusters.* It is widely believed that we are presently in the epoch of cluster formation. According to the anisotropic merger proposed here, clusters of galaxies should form from mergers of smaller subclusters which infall along large-scale filaments. Recent observational work by West, Jones & Forman (1995) has confirmed this picture of cluster formation.

(3) *Dust lanes in cD galaxies.* If cD galaxies are truly prolate, as the model described here requires, then the dynamical arguments outlined in Sections 2 and 4 lead to the prediction that dust and gas disks which are acquired by such galaxies during mergers should settle into planes that are perpendicular to the major axis of the galaxy's mass distribution. Work is in progress to test this prediction.

## 7. Conclusions

Although admittedly speculative in nature, the model proposed here is built upon a foundation of widely accepted theoretical ideas regarding black hole and galaxy formation and the physics of radio sources. What is novel about the present work is that it brings together for the first time a number of seemingly disparate points into a cohesive picture. The most attractive feature of this anisotropic merger model is that it unites two alignment effects that had previously been assumed to be completely independent phenomena.

If once accepts the premise that powerful radio galaxies in the early universe were the precursors of today's cD galaxies, then it does not require a great leap of faith to assume that an anisotropic merger process that can produce cD galaxies with preferred shapes and orientations might also communicate these same anisotropies to the very innermost regions of these galaxies, right down to the black hole central engine. Indeed, given the standard accretion disk model for black hole formation and evolution discussed in Section 4, it would be rather surprising if the spin axes of these objects were completely divorced from their surroundings.

In summary, the model proposed here suggests a remarkable connection between the formation of massive black holes, cD galaxies, and the large-scale structure of the universe.

A more detailed account of this work can be found in West (1994).

**Acknowledgements:** The author was supported by a grant from the NSERC of Canada.

# REFERENCES

Begelman, M.C., Blandford, R.D., and Rees, M.J., 1984, "Theory of Extragalactic Radio Sources", Reviews of Modern Physics 56, 255

Binggeli, B., 1982, "The Shapes and Orientations of Clusters of Galaxies", Astronomy & Astrophysics, 107, 338

Blandford, R.D., and Rees, M.J., 1974, "Twin-Exhaust Model for Double Radio Sources", Monthly Notices of the Royal Astronomical Society, 169, 395

Carlberg, R.G., 1994, "Velocity Bias in Clusters", Astrophysical Journal, 433, 468

Chambers, K.C., Miley, G.K., and van Breugel, W., 1987, "Alignment of Radio and Optical Orientations in High-redshift Radio Galaxies", Nature, 329, 604

Dunlop, J.S., and Peacock, J.A., 1993, "Luminosity Dependence of Optical Activity and Alignments in Radio Galaxies", Monthly Notices of the Royal Astronomical Society, 263, 936

Ellingson, E., Green, R.F., and Yee, H.K.C., 1991, "Clusters of Galaxies Associated with Quasars. II. Galaxy Cluster Dynamics", Astrophysical Journal, 378, 476

Gunn, J.E., 1979, "Feeding the Monster: Gas Discs in Elliptical Galaxies", in Active Galactic Nuclei, Cambridge Univ. Press (Cambridge), p. 213

Harms, R.J., et al., 1994, "HST FOS Spectroscopy of M87: Evidence for a Disk of Ionized Gas Around a Massive Black Hole", Astrophysical Journal, 435, L35

Jaffe, W., Ford, H., Ferraresse, L., van den Bosch, F., and O'Connell, R., 1993, "A Large Nuclear Accretion Disk in the Active galaxy NGC 4261", Nature, 364, 213

Kahn, F., and Woltjer, L., 1959, "Intergalactic Matter and the Galaxy", Astrophysical Journal, 130, 705

Lambas, D.G., Groth, E.J., and Peebles, P.J.E., 1988, "Alignments of Brightest Cluster Galaxies with Large-scale Structures", Astronomical Journal, 95, 996

Lynden-Bell, D., 1969, "Galactic Nuclei as Collapsed Old Quasars", Nature, 223, 690

McCarthy, P.J., van Breugel, W., Spinrad, H., and Djorgovski, S., 1987, "A Correlation Between the Radio and Optical Morphologies of Distant 3CR Radio Galaxies", Astrophysical Journal, 321, L29

Porter, A.C., Schneider, D.P., and Hoessel, J.G., 1991, "CCD Observations of Abell Clusters. V. Isophotometry of 175 Brightest Elliptical Galaxies in Abell Clusters", Astronomical Journal, 101, 1561

Rees, M.J., 1978, "Relativistic Jets and Beams in Radio Galaxies", Nature, 275, 516

Rigler, M.A., Lilly, S.J., Stockton, A., Hammer, F., and Le Fèvre, O., 1992, "Infrared and Optical Morphologies of Distant Radio Galaxies", Astrophysical Journal, 385, 61

Röttgering, H.J.A., West, M.J., Miley, G.K., & Chambers, K.C., 1995, "The Optical Counterparts and the Environments of Ultra-steep-spectrum Radio Sources", Astronomy and Astrophysics, in press

Ryden, B.S., Lauer, T.R., and Postman, M., 1993, "The Shapes of Brightest Cluster Galaxies", Astrophysical Journal, 410, 515

Steiman-Cameron, T.Y., and Durisen, R.H., 1988, "Evolution of Inclined Galactic Gas Disks. I. A Cloud-Fluid Approach", Astrophysical Journal, 325, 26

Tohline, J.E., Simonson, G.F., and Caldwell, N., 1982, "Using Gaseous Disks to Probe the Geometric Structure of Elliptical Galaxies", Astrophysical Journal, 252, 92

Tubbs, A.D., 1980, "The Dynamical Evolution of NGC 5128", Astrophysical Journal, 241, 969

van Kampen, E., and Rhee, G., 1990, "Orientation of Bright Galaxies in Abell Clusters", Astronomy and Astrophysics, 237, 283

West, M.J., 1994, "Anisotropic Mergers at High Redshifts: The Formation of cD Galaxies and Powerful Radio Sources", Monthly Notices of the Royal Astronomical Society, 268, 79

West, M.J., Jones, C., and Forman, W., 1995, "Substructure: Clues to the Formation of Clusters of Galaxies", Astrophysical Journal Letters, 451, L5

West, M.J., & Schombert, J.M., 1996, "Large-scale Alignments of Brightest Cluster Galaxies", to be submitted to the Astrophysical Journal

Yates, M.G., Miller, L., and Peacock, J.A., 1989, "The Cluster Environments of Powerful, High-redshift Radio Galaxies", Monthly Notices of the Royal Astronomical Society, 240, 129

Yee, H.K.C., and Green, R.F., 1987, "Surveys of Fields Around Quasars", Astrophysical Journal, 319, 28

Zel'dovich, Ya.B., and Novikov, I.D., 1971, Relativistic Astrophysics. Vol. I. "Stars and Relativity", Univ. of Chicago Press (Chicago)